# The Main Cognitive Model of Visual Recognition:

# Contour Recognition

Chen Yonghong

Institute of Artificial Intelligence, Dept. of Physics and Electronics, Hanshan Normal University, Chaozhou, Guangdong, P. R. China 521041

**Abstract** In this paper, we will study the following pattern recognition problem: Every pattern is a 3-dimensional graph, its surface can be split up into some regions, every region is composed of the pixels with the approximately same colour value and the approximately same depth value that is distance to eyes, and there may also be some contours, e.g., literal contours, on a surface of every pattern. For this problem we reveal the inherent laws. Moreover, we establish a cognitive model to reflect the essential characteristics of the recognition of this type of patterns. In [1], a coarser model or a basicer one is described. In this paper, some important errors are revised, some key things are added, at last, a complete model is described.

## 1. Introduction

In this paper, we will study the following pattern recognition problem: Every pattern is a 3-dimensional graph, its surface can be split up into some regions, every region is composed of the pixels with the approximately same color value and the approximately same depth value that is distance to eyes, and when recognition is conducted, in order to reduce processing time, middle part of each region must be hollowed out, only leaving contours, so there may also be some contours, e.g., literal contours, on a surface of every pattern. The goal of our study is to discover the inherent laws of the recognition of this type of patterns, and to build up the computing model for revealing the inherent laws and the essential characteristics of the recognition of this type of patterns.

A whole classification forest of patterns, i.e., several classification trees of patterns, in a pattern-base, is composed of every pattern. In other words, by classification, patterns are stored.

The whole process is divided into two parts: recognition process and learning process of patterns. For the first one, by category, patterns are stored in the pattern-base, and are searched with a recognized scene. A pattern, corresponding to a root node of a classification tree in the pattern-base, assumed to be $a$, is turned, i.e., projected, from 3-dimensional into 2-dimensional. Then it is done from 2-dimensional into 1-dimensional, only leaving contours of every region composed of the pixels with the same colour value. At last, 1-dimensional pattern composed of some contours are similarly compressed to a minimum, assumed to be $a'$, i.e., $2^{-k}$ of the original size $(k \in N)$. If further compressed, now there are no distinctions between the various patterns. This makes an amount of information on a piece of image greatly compressed. The processing time of matching is greatly reduced in the subsequent recognition (and learning) procedure. And the storage capacity can be greatly compressed. The recognized object, assumed to be $b$, is turned from 3-dimensional into 2-dimensional, then done from 2-dimensional into 1-dimensional, lastly similarly compressed to a minimum, assumed to be $b'$, i.e., $2^{-l}$ of the original size $(l \in N)$. $b'$ is between $a'$ and the double of $a'$. Then $b'$ is successively amplified, the latter one is the double of the previous one. And every location on every amplified one of $b'$ is searched along left, right, upword and downword directions. The corresponding region at every location on every amplified one is matched with $a'$. (1) If every average matching accuracy rate assumed to be $\bar{\theta}$ is smaller than a certain threshold, it is indicated that the recognition is not successful, and whole procedure is ended. (2) Otherwise, it is indicated that the recognition of the root node is successful. Then every child node is successively (parallelly) treated in the manner similar to the recognition treatment one of the root node. The children nodes unsuccessfully recognized are ignored. If there is a child node successfully recognized, the classification search is continued down next level nodes. Recognition treatment is conducted down the classification tree of patterns in this way, until there is a node, none of whose children nodes is successfully recognized, or that is a leaf node. Then the pattern corresponding to the node shall be required.

Many patterns in the classification tree of patterns may be successively found from the just inputted scene. When every pattern is searched, parallelly searching by category may be adopted. Many patterns in the classification forest of patterns, may be found from the just inputted scene. Based on these, **a procedure that a thing happens** may be recognized from an inputted continuous scene sequence.

As to learning procedure, by the method that a newly inputted pattern and an old one in the pattern-base are recognized from and matched with each other, the position, at which the new pattern may be put, may be found from the pattern-base memorized by category. Then the new pattern is put at the position. Specifically, the newly inputted pattern and every one corresponding to every root node in the classification forest are successively (parallelly) recognized from and matched with each other. (1) If a pattern corresponding to a root node is recognized from the new one, oppositely not established, then the node corresponding to the new pattern is regarded as an offspring of the root one, the new pattern and every one corresponding to every child one of the root node are successively (parallelly) recognized from and matched with each other, recursive process as such is continued, layer by layer. (2) If the new pattern is recognized from a one corresponding to a root node, oppositely not established, then the node corresponding to the new pattern is regarded as the father one of the root one. (3) If any pattern corresponding to any one of root nodes cannot be recognized from the new one, and the latter cannot be recognized from the former, then the node corresponding to the new one is isolatedly put into the classification forest. (4) If the new pattern and a one corresponding to a root node can be recognized from each other, then, in fact, two ones are the same.

When a new pattern is learned, parallel process may be conducted. Many learned new patterns may happen (appear) at the same time. So they may also parallelly be learned. Based on these, **a procedure that a thing happens** may be learned from an inputted continuous pattern sequence.

The rest of this paper is organized as follows: memory and organization of patterns, algorithms of pattern recognition, algorithms of pattern learning. Finally summary and outlook concludes this paper with future work.

## 2. Memory and Organization of Patterns

Memory and organization of various patterns is the basis of pattern recognition. If done well, memory capacity can be greatly compressed, retrieval speed by category can also be greatly increased during pattern recognition, processing time of pattern recognition (matching) can also be significantly reduced, learning speed of a new pattern can also be greatly enhanced, it can help to quickly find the location where the new pattern should be placed.

**Defination 2.1** In a sensory cortex (here it is a visual cortex) in a human brain, there are large areas that are composed of many slender small cuboids (cortical columns[3]). Only pixels in a small local area on surface of a scene are able to be projected and temporarily (not perpetual) stored in each cortical column. A colour value of a neuron at every pixel is activated. The colour values are recorded in polypeptide contained in the neurons. The neurons are known as **the semantic neurons**.

The reason that these large areas in a visual cortex (other sensory cortices are as such) in a human brain are composed of many slender small cortical columns may be: In the evolution procedure, in order to increase storing capacity of these large areas, they must be folded, then surface shape of a scene could not be exactly projected. If they are decomposed into some cortical columns, and a small local area on surface of a scene is able to be exactly projected in every cortical column, and whole surface shape of a scene is defined by positions arranged between cortical columns, the surface shape of a scene is able to be exactly projected.

After leveling, the large areas in a visual cortex in a human brain are large arrays, i.e., large cuboids, composed of slender cortical columns. A scene is projected in a cubic local area in a sensory cortex in a human brain. The length, width, hight of the local area respectively correspond to $X$-axis, $Y$-axis and $Z$-axis in 3-dimensional coordinate system. Coordinates of any pixel on the surface projected by the scene are $(x, y, z)$, ($x \geq 0, y \geq 0, z \geq 0$). The color value of the neuron at this pixel is $<a, b, c, d>$. $a$, $b$ and $c$ represent values of red, green and purple, respectively. They are between 0 and 255. And $d$ represents divergence of light. For example, as to a block of glass, $d$ is 255, and as to glass powder, $d$ is 0. That color value is $<255, 255, 255, 255>$ represents transparent. That color value is $<0,0,0,0>$ rerepresents black. If one point in a primary sensory cortex does not belong to the surface projected by the scene, it's color value is $<255, 255, 255, 255>$, now this pixel (neuron) is at "sleeping" state, i.e., waking one but unresponsive one.

**Definition 2.2** When a graph without background is projected in a cubic local area in a visual cortex in a human brain. It can be 3-dimensional (some 3-dimensional graphic blocks of objects are included), 2-dimensional (the depth information is not temporarily considered, every 2-dimensional graphic block is composed of many pixels of the same color value, many graphic blocks can be included in a 2-dimensional graph), 1-dimensional (the curves (the contours) in 2-dimensional coordinate system or 3-dimensional coordinate system are included), the combination of any 2 kinds of dimension or the one of 3 kinds of dimension. And this graph is used to represent the various features of a type of objects. This graph is called **the contour unit of a type of objects**, abbreviated as **a contour unit**.

A **contour unit** is used to represent an object in a variety of characteristics (attributes). It need not be a complete graph. For example, in order to get **the contour unit** representing the elderly, we only need draw gray hair and wrinkles on a face, do not have to draw a complete face.

**Definition 2.3** On the inner side of a local area in a visual cortex in a human brain, i.e., a $X\_Y$ plane, each pixel (neuron) corresponds to the one (**the semantic neuron**), at waking state, of the same coordinates, in the cortical column above it. The colour value and the depth value of the latter are temporarily (not perpetual) stored in the former, i.e., polypeptide contained in the pixel (neuron). The 2-dimensional representation of the scene or **the contour unit** is composed of these pixels (neurons) on the $X\_Y$ plane. The $X\_Y$ plane is called **the plane of 2-dimensional representation.**

Information input and output of every cortical column are on its inner side. At first, 3-dimensional surface of a scene is projected in retinas in two eyes. Then the retinas is cut into $p$ slices according to the hight. Then through $(km) \times (km)$ neural pathways, information of every slice is parallelly transmitted into the same hight in an array composed of $m \times m$ cortical columns. $k \times k$ pixels can be stored in every cortical column in the array. As such, by sequential $p$ moments, 3-dimensional surface of the scene is projected in the array composed of $m \times m$ cortical columns. then by 1 moment, 3-dimensional representation of the scene is transformed into 2-dimensional representation on **a plane of 2-dimensional representation**. The 2-dimensional representation is outputted through information outputs of these cortical columns. The concrete transformation procedure is in **Algorithm 2.9** in the later section of this paper. As such, these above 2 procedures are uninterruptedly processed in turn.

**Definition 2.4** In a local area, "near" (i.e. linked to) a sensory cortex, in the association cortex in the right hemisphere, there is a 2-dimensional array (a plane area) composed of many neurons. Besides, every node must be respectively linked to 8 ones directly near it by 8 edges (synapses) on this new plane area. A $n \times n$ phalanx is composed of nodes on this new plane area. This new plane area is called **a plane of 1-dimensional representation. A** different type of sensory organ corresponds to a different **plane of 1-dimensional representation.**

As to visual cortices, neurons on **a plane of 2-dimensional representation** directly correspond to neurons on **a plane of 1-dimensional representation**. They are linked by neural pathways. Their connective manners are as follows. A coordinate system similar to one on **the plane of 2-dimensional representation** is set up on **the plane of 1-dimensional representation**. Two coordinate systems are parallel. And one vertically corresponds to another by logic. Every node (neuron) $(x, y)$ on **the plane of 1-dimensional representation** is linked to the node (pixel) $(x, y)$ on **the plane**

**of 2-dimensional representation** by an edge. At the same time, respectively linked to 8 pixels directly near the latter by 8 edges.

A sensory cortex is either in the right hemisphere in a human brain, or in the left hemisphere in a human brain. **A plane of 1-dimensional representation** is subjected to the association cortex in the right hemisphere in a human brain. Sensory cortices (A visual cortex is directly substituded with **A plane of 2-dimensional representation**.), not only in the right hemisphere, but also in the left hemisphere, in a human brain, are linked to **a plane of 1-dimensional representation** subjected to the association cortex in the right hemisphere through neural pathways. Here there is an important explanation that, as to the neural pathways linking **a planes of 2-dimensional representation** subjected to **a** sensory cortex in the left hemisphere in a human brain to **a plane of 1-dimensional representation** subjected to the association cortex in the right hemisphere, the right hemisphere and the left one are linked through them and corpus callosum together. There are other neural pathways related to auditory sense, somatic sensation and so on. Function of corpus callosum will be described in the later section of the paper.

Besides, visual cortices are at pillow pages in the right hemisphere and the left one in a human brain. The nodes of the same coordinates on **Planes of 2-dimensional representation** subjected to visual cortices in the right hemisphere and the left one in a human brain are linked to the same node on **a plane of 1-dimensional representation** subjected to the association cortex in the right hemisphere through neural pathways. In other words, the methods that two **planes of 2-dimensional representation** are linked to **the plane of 1-dimensional representation** are the same. So, the scene everyone sees by his right eye is the same as the one he does by his left one.

**Definition 2.5** In a local area near **a plane of 1-dimensional representation** in the association cortex in the right hemisphere, there is a 2-dimensional array (plane area) composed of many neurons. The neurons on it directly correspond to the neurons on **the plane of 1-dimensional representation.** They are linked by edges (synapses). Synaptic connective manner is as follows. A coordinate system similar to one on **the plane of 1-dimensional representation** is set up on this new plane area. Two coordinate systems are parallel. And one vertically corresponds to another by logic. Every node (neuron) on this new plane area is vertically pointed at (linked to) a corresponding one on **the plane of 1-dimensional representation** by an edge. Suppose that coordinates of a node on this new plane are $(x_0, y_0)$, and the coordinates of the corresponding one on **the plane of 1-dimensional representation** are $(u, v)$, then if coordinates of a node on this new plane are $(x_1, y_1)$, then the coordinates of the corresponding one on **the plane of 1-dimensional representation** are $(u + x_1 - x_0, v + y_1 - y_0)$. If there are $(2l+1) \times (2l+1)$ nodes on **the plane of**

**1-dimensional representation**, they formed a phalanx, this phalanx is divided into some small phalanxes, every small phalanx is composed of $3 \times 3$ nodes, then every node $(x, y)$ on this new plane area, must be linked to the centre node $(2x+1, 2y+1)$ on the corresponding small phalanx by an edge, at the same time, respectively linked to 8 nodes directly near this centre node by 8 edges. Besides, every node must be respectively linked to 8 ones directly near it by 8 edges on this new plane area. The nodes on this new plane area formed a $l \times l$ phalanx. Again, other plane area are set up on this new plane area. And so on, many plane areas and corresponding phalanxes are set up in the association cortex in the right hemisphere. These plane areas are called **conversion planes**.

**Definition 2.6** On **a plane of 1-dimensional representation** or **a conversion plane**, a group of contours that represents a type of objects is limited in a square frame. Every pixel in the group of contours is represented as a neuron. The group of contours that represents every type of objects, and the corresponding square frame are compressed synchronously. Every time they are compressed to the half of the original size. There is edge-length of a square frame that is assumed to be $\omega$ pixels. If they continue to be compressed synchronously, and the edge-length of the corresponding square frame is smaller than $\omega$ pixels, at this time, every type of objects cannot be recognized from the group of contours. If a graph is a group of contours on a plane, and the group of contours is limited in a rectangle whose width is $w_1$ pixels, and whose length is $w_2$ pixels, $\omega \leq w_1 \leq w_2 < 2\omega$, there is a square frame whose edge-length is $w_2$, and in which the group of contours is limited. The center point of this square frame coincides with the center point of the above rectangle. The four edges of this square frame are respectively parallel to the four edges of the above rectangle. The graph is used to directly represent the features of a type of objects. The graph is called **the contour feature of a type of objects**, abbreviated as **a contour feature.**

**Definition 2.7** In the association cortex in the right hemisphere, synapses of every neuron in corpus callosum[6] are either all absent, or respectively pointed at (linked to) different pixels in **a contour feature** (as to the visual systems), one to one. As to every pixel, there is a synapse pointed at it. A neuron in corpus callosum is known as **a position neuron**. Every **contour feature** directly corresponds to **a position neuron**. **Position neurons** in corpus callosum formed a 2-dimensional array. The position of every **position neuron** is represented as coordinates $(x, y), (x \geq 0, y \geq 0)$. They are called **position coordinates**.

**Definition 2.8** On **a plane of 1-dimensional representation** or **a conversion plane** in the association cortex in the right hemisphere, $P_1, P_2 ... P_n$ are recorded in every neuron $a$. $P_j = [a_j, b_j, c_j, d_j; z_j; x_j, y_j]$, $(1 \leq j \leq n)$. $P_j$ represents basic information of a pixel on 1-dimensional representation of a scene or **a contour unit**, i.e., some contours representing a scene or **a contour unit**. $<a_j, b_j, c_j, d_j>$ represents the colour information of the pixel. It is called **a relative colour value**. $z_j$ represents the depth information of the pixel. It is called **a relative depth value**. **The position coordinates** of the corresponding **position neuron** in corpus callosum are $(x_j, y_j)$. $P_1, P_2...P_n$ are recorded in polypeptide contained in neuron $a$. And they are respectively recorded at different time. $P_j$, $(1 \leq j \leq n)$, is known as **a position feature value.** $[a_j, b_j, c_j, d_j; z_j]$, $(1 \leq j \leq n)$, is known as **a feature value**.

**Parallel Algorithm 2.9** 3-dimensional representation of an inputted scene or **a contour unit** is transformed into 2-dimensional representation.

In a local area in a primary sensory cortex, where there is 3-dimensional representation of a scene or **a contour unit**, the coordinate of any pixel (now it is at waking state) on the surface of the scene or **the contour unit** are $(x, y, z)$. It's color value is $<a, b, c, d>$, then this pixel is projected at $(x, y)$ on **the plane of 2-dimensional representation** and turned into a new pixel (neuron) (now it is at waking state). The color value $<a, b, c, d>$ and the depth value $z$ are temporarily recorded in this neuron. If any spot $(u, v, w)$ （$0 \leq w \leq L$） in the primary sensory cortex is not subjected to the surface of the scene or **the contour unit**, then the color value and the depth value of the pixel $(u, v)$ on the corresponding plane $X\_Y$ are $<255, 255, 255, 255>$ and $L$, now this pixel (neuron) is at "sleeping" state, i.e., waking one but unresponsive one. $L$ is a constant, it represents the greatest depth value of objects, it actually is the greatest depth value of a cortical column.

Now we discuss the following algorithm. Our eyes are unresponsive to light colour, but sensitive to heavy colour, especially black. They are unresponsive to light with low divergence, but sensitive to light with high divergence. They are unresponsive to a far place, but sensitive to a near place. And a distance from an object to eyes is not the depth (the hight) of the object itself. Based on the above fact, we can approximately find the following conclusions. If a colour value and a depth value of any pixel $(x, y)$ on **the plane of 2-dimensional representation** of a scene or **a contour unit** are $<a, b, c, d>$ and $z$, respectively, and if the node $(x, y)$ on **the plane of 1-dimensional representation** is activated, then **the feature value** of the latter is $[255-a, 255-b, 255-c, 255-d; L-z]$. Meanwhile, **the position coordinates** in the corresponding **position feature value** are initialized into $(-1, -1)$. That they are negative is because the coordinates greater than or equal to 0 are all valid.

**Parallel Algorithm 2.10** 2-dimensional representation of an inputted scene or **a contour unit** is transformed into 1-dimensional representation.

(1) if a colour value and a depth value of any pixel $(x, y)$ on **the plane of 2-dimensional representation** are $<a, b, c, d>$ and $z$, respectively, and the pixel $(x, y)$ and other 8 pixels directly near it conform to the following conditions:

((1)) their colour values are approximately equal, i.e., the difference of any two colour values is smaller than a threshold, or the condition is not founded, but $<255-a, 255-b, 255-c, 255-d> \approx <0, 0, 0, 0>$,

((2)) suppose that any 3 pixels arranged in a line (There are 8 groups) are A, B and C, successively, and their depth values are $z_A$, $z_B$ and $z_C$, successively, then $|(z_A - z_B) - (z_B - z_C)| = |z_A + z_C - 2 \cdot z_B|$ is smaller than a threshold, or the condition is not founded, but $L - z \approx 0$, then **the feature value** of the node $(x, y)$ on **the plane of 1-dimensional representation** is $[0, 0, 0, 0; 0]$. The node is at sleeping state.

(2) If these 9 pixels do not conform to one of the above two conditions, then **the feature value** of the node $(x, y)$ on **the plane of 1-dimensional representation** is $[255-a, 255-b, 255-c, 255-d; L-z]$.

On **the plane of 1-dimensional representation**, the nodes, whose **feature values** are not $[0,0,0,0;0]$, form several contours. They are the 1-dimensional representation of the scene or **the contour unit**.

From the above algorithm we can find the following conclusions. (1) The junction spot of two pieces of regions, whose colour values are greatlier different from each other, whose colour values are heavier and possess higher divergence, and whose depth values conform to the other condition, on **the plane of 2-dimensional representation** of an inputted scene or a recognized **contour unit**, can be transformed into two closely adjacent curves on **the plane of 1-dimensional representation**. There is not a pixel whose **feature value** is $[0,0,0,0;0]$ between two curves. **The relative colour value** of every pixel on any one of two curves is the same. If **the relative colour value** is $<a,b,c,d>$, $<255-a, 255-b, 255-c, 255-d>$ represents the colour value of the region circled by this curve on **the plane of 2-dimensional representation**. On **the plane of 2-dimensional representation**, the segments of the contour around the region are horizontal lines, or vertical lines, or slashs whose angle are 45°. On two closely adjacent curves on **the plane of 1-dimensional representation**, the original horizontal lines on **the plane of 2-dimensional representation** now are turned into horizontal lines, and the original vertical lines are done into vertical lines, and the original slashs whose angle are 45° are done into the curves which are composed of the shortest horizontal and vertical segment, similar to ladders. Anyway, the curves are composed of horizontal segments and vertical segments. (2) Besides, similar to (1), the junction spot of 2 pieces of areas, whose depth values are suddenly changed, whose depth values are smaller, and whose colour values conform to the other condition, on **the plane of 2-dimensional representation** of an inputted scene or a recognized **contour unit**, can be transformed into one curve on **the plane of 1-dimensional representation.** (3) At last, as to a black curve on background, whose colour is whiter or lighter, whose depth value are not suddenly changed, e.g., black characters on a piece of white paper put flatly, or as to a white curve on background, whose colour is blacker or heavier, there is the following situation: After any pixel on a curve hading existed long before on **the plane of 2-dimensional representation** of an inputted scene or a recognized **contour unit**, and the other 8 pixels directly near it are transformed, their **relative colour values** are not $<0,0,0,0>$ on **the plane of 1-dimensional representation**. In other words, an 1-dimensional curve hading existed long before on **the plane of 2-dimensional representation** can be transformed into 3 closely adjacent curves on **the plane of 1-dimensional representation**.

Overall, the group of contours on **the plane of 1-dimensional representation** is composed of a curve, 2 closely adjacent ones, or 3 ones.

**Parallel Algorithm 2.11** By conversion between 2 adjacent **conversion plane**s, 1-dimensional representation or compressed one of a scene or **a contour unit** is compressed to the half of the original size.

Between two adjacent **conversion plane**s, or between **a plane of 1-dimensional representation** and a adjacent **conversion plane** (The followings are as such), we suppose that there are $(2l+1)\times(2l+1)$ nodes on the lower plane, and these nodes form a phalanx, and the phalanx is divided into some small phalanxes, and every small phalanx is composed of $3\times 3$ nodes. The following operations are executed for every small phalanx. We suppose that a center node of a small phalanx is $(2x+1, 2y+1)$, then a corresponding one is $(x, y)$ on the above plane. Now the nodes whose **feature value** are not $[0,0,0,0;0]$ are all at waking state. The ones whose **feature value** are $[0,0,0,0;0]$ are all at sleeping state.

(1) If **the feature value** of the node $(2x+1, 2y+1)$ on the lower plane is $[0,0,0,0;0]$, and the ones of the other 8 nodes directly near it are also $[0,0,0,0;0]$, then the one of the corresponding node $(x, y)$ on the above plane is $[0,0,0,0;0]$.

(2) If **the feature value** of the node $(2x+1, 2y+1)$ on the lower plane is not $[0,0,0,0;0]$, suppose that it is $[a_1, b_1, c_1, d_1; z_1]$, then the one of the corresponding node $(x, y)$ on the above plane is $[a_1, b_1, c_1, d_1; z_1/2]$.

(3) If **the feature value** of the node $(2x+1, 2y+1)$ on the lower plane is $[0,0,0,0;0]$, and the ones of the other 8 nodes directly near it are not $[0,0,0,0;0]$ at all, then the one of the corresponding node $(x, y)$ on the above plane is not $[0,0,0,0;0]$. From the nodes, in every phalanx on the lower plane, whose **feature values** are not $[0,0,0,0;0]$, the one that is the nearest to the corresponding one $(x, y)$ on the above **conversion plane**, i.e., that is the nearest to the center one $(2x+1, 2y+1)$ of the phalanx, is selected, to be used to activate the corresponding one $(x, y)$ on the above **conversion plane**. Suppose that its **feature value** is $[a_2, b_2, c_2, d_2; z_2]$, then the one of the corresponding node $(x, y)$ on the above **conversion plane** is $[a_2, b_2, c_2, d_2; z_2/2]$.

And so on, the operations are executed for $k$ times by the above method. The 1-dimensional representation (The contours) of the inputted scene or **the contour unit** can be compressed to $1/2^k$ of the original size, $k \in N$.

In the compressed procedure, **the position coordinates** in the corresponding **position feature values** are initialized into $(-1, -1)$. That they are negative is because the coordinates greater than or equal to 0 are all valid.

From the algorithm we can find the following conclusions. As to (3), if the nodes that is the nearest to the corresponding one $(x, y)$ on the above **conversion plane** are more than one, then the following method is processed. ((1)) If there are ones in center nodes at 4 edges in every phalanx, whose **feature values** are not $[0,0,0,0;0]$, then which is selected from them, to be used to activate one on the above **conversion plane**, is fuzzy and not accurate. ((2)) Otherwise, if there are ones in nodes at 4 angles in every phalanx, whose **feature values** are not $[0,0,0,0;0]$, then which is selected from them, to be used to activate one on the above **conversion plane**, is fuzzy and not accurate. ((3)) The fuzzy and not accurate characteristics on the above two hands are inherent and insurmountable.

If **a contour unit** representing a type of objects contains 3-dimensional or 2-dimensional graphic sections, the 3-dimensional graphic sections are transformed into 2-dimensional ones, the 2-dimensional ones are done into 1-dimensional ones. The 1-dimensional contours representing the characteristic of the type of objects are left. If a colour value and a depth value of a pixel in **the contour unit** are $<a, b, c, d>$ and $z$, respectively, **the relative colour value** and **the relative depth value** of the corresponding one in the contours are $<255-a, 255-b, 255-c, 255-d>$ and $L-z$, respectively. We suppose that the 1-dimensional contours are limited in a square frame whose edge-lengths are $p$ pixels.

At last, the square frame and the contours are synchronously compressed to $1/2^k$ of the original sizes until $\omega \leq p/2^k < 2\omega, k \in N$, in which $\omega$ appears in **Definition 2.6**. As so, **a contour unit** is transformed into a corresponding **contour feature**. If every **contour feature** is compared with a corresponding **contour unit**, the graphic sections more than 1-dimensional in **the contour unit** are transformed into the 1-dimensional contours, then the contours are similarly compressed to minima. This makes the amount of the information of the formed **contour feature** greatly compressed, and the time complexity of the matching during subsequent recognition and learning greatly decreased, and the storage (memory) capacity greatly compressed.

**Parallel Algorithm 2.12** By conversion between 2 adjacent **conversion planes**, compressed 1-dimensional representation of an inputted scene or a newly learned **contour unit** is enlarged to the double of the original size.

The algorithm is the reverse process of **Algorithm 2.11**. Because the scene or **the contour unit** is put into a brain right now, the neurons on every compressed or not 1-dimensional representation of the scene or **the contour unit**, on **a plane of 1-dimensional representation** and **conversion planes,** are all at waking state.

**Parallel Algorithm 2.13** By conversion between 2 adjacent **conversion planes**, 1-dimensional representation of **a contour feature** or its $2^k$ multiple is enlarged to the double of the original size. **Position coordinates** of a corresponding **position neuron** in corpus callosum, which are just (permanently) recorded in the neurons at waking state on the above plane, are sent down, permanently recorded in the neurons at waking state on the lower plane.

The algorithm is similar to the reverse procedure of **Algorithm 2.11**. If the node $(x, y)$ on the above plane are at waking state, and the new **position feature value** $[a,b,c,d;z;u,v]$ is just (permanently) recorded in polypeptide contained in the neuron, then **the position coordinates** (u，v) of the corresponding **position neuron** in corpus callosum，which are just (permanently) recorded in the neurons, are sent down, every node at waking state in 9 nodes centering around the node $(2x+1, 2y+1)$ on the lower plane puts its **feature value** and (u，v）together, a new **position feature value** is formed, and it is permanently recorded in polypeptide contained in the node (neuron).

The algorithm is used to the procedure of pattern learning.

**Parallel Algorithm 2.14** By conversion between 2 adjacent **conversion planes**, 1-dimensional representation of **a contour feature** or its $2^k$ multiple wakes the one of the enlarged double of the original size on the lower plane.

The algorithm is similar to the reverse procedure of **Algorithm 2.11**.

If the node $(x, y)$ on the above plane is at waking state, its activated **position feature value** is $[a,b,c,d;z;u,v]$, then the nodes in 9 nodes centering around the node $(2x+1, 2y+1)$ on the lower plane, which contain **the position feature values** whose formats are $[\_,\_,\_,\_;\_;u,v]$, are activated, are at waking state, **the position feature values**, whose formats are $[\_,\_,\_,\_;\_;u,v]$, in activated nodes, are also simultaneously activated.

The algorithm is used to the procedure of pattern recognition.

It is noted that the above 4 algorithms have nothing to do with **a plane of 2-dimensional representation**, and only **feature values** are recorded in neurons on 2-dimensional representation of an inputted scene or **a contour unit** corresponding to **a contour feature**, overall **position feature values** are not recorded in them.

**Definition 2.15** Either every **position neuron** in corpus callosum has not a synapse in the association cortex in the right hemisphere, or its synapses are only pointed at pixels on **a contour feature** (as to the visual systems). Still, everyone of the latter **position neurons** is linked to a neuron in the association cortex in the left hemisphere by a synapse. Then every **contour feature** in the right hemisphere corresponds to a neuron in the association cortex in the left hemisphere. We suppose that the neurons conforming to the condition, in the association cortex in the left hemisphere, are $a_1, a_2 \ldots a_k$. These $k$ neurons are linked to each other. Several tree structures of space classification are formed. A father neuron is linked to a son neuron by a synapse. And the synapse is from the father neuron to the son neuron. **The contour feature**, which the father neuron corresponds to, can be found from **the contour feature** which a son neuron corresponds to. Several tree structures conforming to the above condition are known as **a pattern classification forest**. If there only is a tree structure, it is known as **a pattern classification tree**.

As to every **contour feature** in the right hemisphere, there is a neuron corresponding to it in the association cortex in the left hemisphere. The neuron represents a type of objects. The feature of the type of objects can be directly represented as **the contour feature,** or the corresponding **contour unit**. In fact, the neuron represents a pattern in a pattern-base. Besides, every neuron conforming to the condition in the left hemisphere is linked to each other by some synapses. **A pattern classification forest** is formed. It represents the static space classification of all **contour units**. Searching **a pattern classification forest** by classification, and finding some neurons, corresponding to some **contour units** (some **contour features**) coming from an inputted scene from outside, in the association cortex in the left hemisphere, are the most basic procedures of contour recognition.

**Defination 2.16** Hippocampus[7] is a linear structure composed of neurons. The time sequence a person experienced before, $T_0$, $T_0+\Delta T$, $T_0+2\Delta T$,…, $T_0+n\cdot\Delta T$, is recorded in hippocampus, in which $T_0$ is the beginning time, and $\Delta T$ is solid time interval, and $T_0+n\cdot\Delta T$ is the present time. Every segment of the time sequence is recorded in multipeptide contained in a neuron. These neurons are called **time neurons**. **The time neuron** which the present time $T_0+n\cdot\Delta T$ is recorded in is called **the present neuron**. In the linear structure, the previous times are recorded in **the time neurons** in front of **the present neuron**. None of times is recorded in the ones in the back. At every $\Delta T$, **the present neuron** will produce and remember a signal representing a new time. When the times which are recorded in multipeptide in **the present neuron** overflow, the next **time neuron** in the linear structure must be regarded as **the present neuron**.

**Definition 2.17** Several times $T_{i1}$, $T_{i2}…T_{il_i}$ are recorded in every neuron $a_i$ ($1\leq i\leq k$) in **a pattern classification tree (forest)** in the association cortex in the left hemisphere in a human brain. These times are recorded in multipeptide contained in $a_i$. Where $T_{i1}<T_{i2}<…<T_{il_i}$. The neuron is linked to **the time neuron** where the time $T_{i1}$ is recorded in hippocampus by a synapse. **The time neuron is the present neuron** that appeared when **a position neuron** in corpus callosum was just linked to the neuron $a_i$ by a synapse. A time $T_{ij}$ recorded in every neuron $a_i$ directly corresponds to the time $T_{ij}$ recorded in **a time neuron** in hippocampus. Every **contour unit** that appears only at a time recorded in **a time neuron** in hippocampus can be recorded in a human brain. The sequence, arranged with every **contour unit**, that appears in a segment of time sequence, recorded in **time neurons** in hippocampus, $T,T+\Delta T,T+2\Delta T…T+l\cdot\Delta T$, is known as **a procedure that a thing happens**. Sometimes we simply use $<T,T+l\cdot\Delta T>$ to represent it.

A time $T_{ij}$ recorded in every neuron $a_i$ represents the time at which a corresponding **contour unit** appears one time. If any time $T_{ij}$ appears not only in a neuron $a_i$, but also in the other neuron, this means that many **contour unit** simultaneously appear at the time $T_{ij}$. So many **contour units** may simultaneously appear at some time in **a procedure that a thing happens**. If any time $T_{ij}$ only appears in a neuron $a_i$, does not appear in the other neuron, this means that only **a contour unit** appears at this time.

Thus, the storage and the arrangement of every pattern, i.e., **a contour unit**, used to the pattern recognition, have been completely described in the paper.

## 3. Algorithms of Pattern Recognition

The main task of pattern recognition towards contours is to recognize **a contour unit**, i.e., a type of objects stored in a pattern-base, from an inputted scene, even to recognize **a procedure that a thing happens**, from an inputted successive sequence of scenes. So at first, we introduce the child algorithm recognizing a required **contour feature** from a spot on compressed or not 1-dimensional representation of a scene or **a contour unit** (used to pattern learning).

**Parallel Algorithm 3.1** As to every pixel on a required **contour feature**, its matched pixel (if it exits) is got from a contour in a recognized area on compressed or not 1-dimensional representation of an inputted scene or a recognized **contour unit** (used to pattern learning). If the required **contour feature** and the contours are on the same plane.

A required **contour feature** centers around the point done around by the square frame that limits it. The center point of the required **contour feature** coincides with one of a recognized area on compressed or not 1-dimensional representation of a recognized scene or a recognized **contour unit**. We assume that it is $(x_0, y_0)$. (If $(x_0, y_0)$ is on a contour, i.e., its **feature value** is not $[0,0,0,0;0]$, the point is appropriately deviated. If $(x_0, y_0)$ is not a pixel, i.e., $x_0$ and $y_0$ are not integers, the point is appropriately deviated). As to every pixel (at waking state at this time) on the required **contour feature**, the following treatment is simultaneously executed.

Because the 1-dimensional representation of the inputted scene or the recognized **contour unit** is compressed to the half of the original size every time, if the required **contour feature** can be recognized from a recognized area on the compressed or not 1-dimensional representation of the recognized scene or the recognized **contour unit**, the group of the recognized contours assumed to be $C$ is similar to the required **contour feature** assumed to be $B$, and on the average, $C$ is between $B$ and the double of $B$, assumed to be a $\mu$ multiple of $B$, in which $1 \leq \mu < 2$. We suppose that the distance between anyone of the pixels on $B$, not $(x_0, y_0)$, and $(x_0, y_0)$ is $l_1$, the distance between the matched pixel on $C$ and $(x_0, y_0)$ is $l_2$. The matched pixel may be deviated. But its deviation must conform to the following condition: $0 < l_2 < 2\mu l_1$, i.e., $0 < l_2/l_1 < 2\mu < 4$. So we suppose that $(x_1, y_1)$ is anyone of the pixels on the requireed **contour feature,** but is not $(x_0, y_0)$. The pixel, matched with $(x_1, y_1)$, on the compressed or not 1-dimensional representation of the recognized scene or the recognized **contour unit,** must conform to the following 5 conditions. We suppose that the pixel is $(<x_2>, <y_2>)$. In fact, it is a point $(x_2, y_2)$ on the ray (conforming to (3) and (4)). The coordinates of the pixel the closeliest near $(x_2, y_2)$ are $(<x_2>, <y_2>)$.

$$0 < \frac{y - y_0}{y_1 - y_0} < 4 \qquad \text{(when } y_1 \neq y_0 \text{)} \qquad (1)$$

$$y = y_0 \qquad \text{(when } y_1 = y_0 \text{)} \qquad (2)$$

$$0 < \frac{x - x_0}{x_1 - x_0} < 4 \qquad \text{(when } x_1 \neq x_0 \text{)} \qquad (3)$$

$$x = x_0 \qquad \text{(when } x_1 = x_0 \text{)} \qquad (4)$$

$$\frac{y - y_0}{x - x_0} = \frac{y_1 - y_0}{x_1 - x_0} \qquad \text{(when } x_1 \neq x_0 \text{)} \qquad (5)$$

The ray from $(x_0, y_0)$ through $(x_1, y_1)$, i.e., conforming to (5) or (4), may approximately pass through, i.e., be closely near, many not adjacent pixels on the required **contour feature**, i.e., there are many points of intersection between the ray and the curve. As such, a segment on the ray from $(x_0, y_0)$ through $(x_1, y_1)$, i.e., conforming to (1),(2),(3),(4) and (5), may approximately pass through, i.e., be closely near, many not adjacent pixels on the compressed or not 1-dimensional representation of the recognized scene or the recognized **contour unit**. The two groups of pixels successively correspond to each other, one to one, from $(x_0, y_0)$ along the ray. So they may be synchronously matched with each other as much as possible. The matching of two groups of contours is the parallel matching of the corresponding pixels, i.e., the matching which are synchronous as much as possible.

$(<x_2>, <y_2>)$ is possibly matched with many pixels on the required **contour feature**. The pixel $(<x_2>, <y_2>)$ matched with $(x_1, y_1)$ possibly does not exist. If $(<x_2>, <y_2>)$ exists, the comparision (recognition) of the location coordinates, **the relative colour values** and **the relative depth values** of the 1-dimensional contours may be conducted.

**Definition 3.2** Suppose that a required **contour feature** and a compressed or not 1-dimensional representation of an inputted scene or a recognized **contour unit** are on the same plane. **The contour feature** is going to be recognized from a location on a compressed multiple of the 1-dimensional representation. As to a pixel $(x_1, y_1)$ of the former, not $(x_0, y_0)$, and a corresponding one $(<x_2>, <y_2>)$ of the latter, their matching accuracy-rate is defined as their **recognition accuracy-rate.** The average value of all of **recognition accuracy-rate**s is defined as **the average recognition accuracy-rate .**

**Algorithm 3.3 The recognition accuracy-rate**, that **a contour feature** is going to be recognized from a location on a compressed multiple of 1-dimensional representation of an inputted scene or a recognized **contour unit**, is got as follows.

By **Algorithm 3.1**, every pixel (neuron) on the required **contour feature,** $(x_{11}, y_{11})$ ... $(x_{1r}, y_{1r})$, $(x_{1,r+1}, y_{1,r+1})$ ... $(x_{1m}, y_{1m})$ (as to every $1 \le j \le m$, $(x_{1j}, y_{1j}) \ne (x_0, y_0)$ ) is searched. A pixel (neuron) $(<x_{2i}>, <y_{2i}>)$ $(1 \le i \le r)$ matched with everyone $(x_{1i}, y_{1i})$ is computed (found out) on compressed or not 1-dimensional representation of an inputted scene or a recognized **contour unit**.

Every pixel $(<x_{2i}>, <y_{2i}>)$ possibly repeats to appear. A pixel (neuron) matched with everyone $(x_{1i}, y_{1i})$ $(r+1 \leq i \leq m)$ does not exist.

(1) The group of the recognized contours known as $C$ is similar to the required **contour feature** known as $B$. On the average, $C$ is between $B$ and the double of $B$, assumed to be a $\mu$ multiple of $B$, in which $1 \leq \mu < 2$. $\mu$ is the following arithmatic average value.

$$\mu_i = \begin{cases} \dfrac{<x_{2i}> - x_0}{x_{1i} - x_0} & \text{if } x_{1i} \neq x_0 \\ \dfrac{<y_{2i}> - y_0}{y_{1i} - y_0} & \text{if } y_{1i} \neq y_0 \end{cases}$$

$$\mu = \frac{\sum_{i=1}^{r} \mu_i}{r}$$

(2) A location deviation between every pixel $(x_{1i}, y_{1i})$ and a corresponding one $(<x_{2i}>, <y_{2i}>)$ $(1 \leq i \leq r)$ is computed as follows:

$$\alpha_i = \left|\vec{\alpha_i}\right| = \frac{\left|\dfrac{<y_{2i}> - y_0}{y_{1i} - y_0} - \mu\right|}{\mu}$$

$\vec{\alpha_i}$ $(1 \leq i \leq r)$ represents a location deviation during the 1-dimensional contours recognition. Its direction is from every pixel $(x_{1i}, y_{1i})$ to a corresponding one $(<x_{2i}>, <y_{2i}>)$ or from a corresponding one $(<x_{2i}>, <y_{2i}>)$ to every pixel $(x_{1i}, y_{1i})$. $0 \leq \alpha_i \leq 3$.

(3) A colour deviation between every pixel $(x_{1i}, y_{1i})$ and a corresponding one $(<x_{2i}>, <y_{2i}>)$ is computed as follows:

We suppose that **the relative colour values** in **the feature values** of the two pixels (neurons) are $<a_{i1}, a_{i2}, a_{i3}, a_{i4}>$ and $<b_{i1}, b_{i2}, b_{i3}, b_{i4}>$, respectively. Then the colour deviation $\beta_i$ is given by

$$\beta_i = \frac{|b_{i1} - a_{i1}| + |b_{i2} - a_{i2}| + |b_{i3} - a_{i3}| + |b_{i4} - a_{i4}|}{4 \times 255}$$

where $0 \leq \beta_i \leq 1$.

(4) In the following, a depth deviation of every pixel $(x_{1i}, y_{1i})$ and a corresponding one $(<x_{2i}>, <y_{2i}>)$ is computing:

We suppose that **the relative depth value** of the pixel $(x_{1i}, y_{1i})$ is $z_{1i}$, and $z_{1i}$ is included in **the feature value** of the neuron $(x_{1i}, y_{1i})$, and the one of the corresponding pixel $(<x_{2i}>, <y_{2i}>)$ is $z_{2i}$, and $z_{2i}$ is included in **the feature value** of the neuron $(<x_{2i}>, <y_{2i}>)$. And suppose that the center point of the required **contour feature** coincides with the one of the recognized area limiting the compressed or not 1-dimensional representation of the recognized scene or the recognized **contour unit**. We assume that it is $(x_0, y_0)$, and **the relative depth value** of the former center point is $z_0$, and **the relative depth value** of the latter one is $z_0'$. Then the depth deviation $\beta_i'$ is:

$$\beta_i' = \frac{|(z_{2i} - z_0') - \mu(z_{1i} - z_0)|}{L \cdot 2^{-k}}$$

where $0 \leq \beta_i' \leq 1$. $k$ is the compressed multiple of the recognized scene or the recognized **graphic unit**. $k \geq 0$. $L$ is a constant. It represents the greatest depth value of objects. It actually is the greatest depth value of **a cortical column**. $L \cdot 2^{-k}$ represents the greatest depth value of similar objects, as to the plane, where the compressed 1-dimensional representation of the recognized scene or the recognized **graphic unit** is stored. It is a constant, as to the plane. In the following, computing **the recognition accuracy-rate,** that **a contour feature** is going to be recognized from a location on a compressed multiple of 1-dimensional representation of an inputted scene or a recognized **contour unit,** is computed.

(5) In the following, computing **a recognition accuracy-rate**[4] that **a contour feature** is going to be recognized from a location on a compressed multiple of 1-dimensional representation of an inputted scene or a recognized **contour unit**.

It is not difficult to see that a accuracy-rate in 1-dimensional contour recognition must conform to Normal Distribution is reasonable, a accuracy-rate in color value recognition is as such. Suppose that, as to every pixel $(x_{1i}, y_{1i})$ $(1 \leq i \leq r)$, a corresponding one, at a location on a compressed multiple of 1-dimensional representation of an inputted scene or a recognized **contour unit**, is $(<x_{2i}>, <y_{2i}>)$. Their **recognition accuracy-rate** is:

$$\theta_i = e^{-C_1 \alpha_i^2 - C_2 \beta_i - C_3 \beta_i'}$$

where $C_1, C_2, C_3$ is a positive constant

In order to compute this formula, we directly recall series expansion. Obviously, two neurons in a human brain cannot conduct so-complicated and so-artificial computing. Only the simplest and the naturalest fuzzy process methods of nerve excitability, e.g., "comprehension", "experience" and "feeling", are adopted.

By parallel process method, **the average recognition accuracy-rate** of the whole required **contour feature** is:

$$\bar{\theta} = \frac{\sum_{i=1}^{r} e^{-C_1 \alpha_i^2 - C_2 \beta_i - C_3 \beta_i'}}{m}$$

Only if $\bar{\theta}$ is greater than or equal to a threshold $C$, is the recognition considered to be successful.

**Theorem 3.4 A contour feature** is going to be recognized from a location on a compressed multiple of 1-dimensional representation of an inputted scene or a recognized **contour unit**. As to every pixel $(x_{1i}, y_{1i})$ $(1 \leq i \leq r)$ of the former, not

$(x_0, y_0)$, and a corresponding one $(<x_{2i}>, <y_{2i}>)$ of the latter, their **recognition accuracy-rate** is

$$\theta_i = e^{-C_1\alpha_i^2 - C_2\beta_i - C_3\beta_i'}$$

where $C_1, C_2, C_3$ is a positive constant

**Proof**

At first, we consider **a recognition accuracy-rate** $\theta_{i1}$ that is related to a location deviation. As to every pixel $(x_{1i}, y_{1i})$ $(1 \leq i \leq r)$, not $(x_0, y_0)$, and a corresponding one $(<x_{2i}>, <y_{2i}>)$, their location deviation is $\vec{\alpha}_i$. It is not difficult to see that **the recognition accuracy-rate** $\theta_{i1}$ is closely related to the circle (in fact, the area of the circle) whose center is the pixel $(x_{1i}, y_{1i})$ and whose radius is $\alpha_i$, i.e.,

$$\theta_{i1} = f(\pi\alpha_i^2)$$

Besides, the components of $\vec{\alpha}_i$ along $X$-axis and $Y$-axis are $\vec{\alpha}_{ix}$ and $\vec{\alpha}_{iy}$, respectively, i.e., $\vec{\alpha}_i = \vec{\alpha}_{ix} + \vec{\alpha}_{iy}$. In other words, the recognition accuracy-rate $\theta_{i1}$ is the multiplication of $\theta_{ix1} = f(\pi\alpha_{ix}^2)$, brought by the circle (in fact, the area of the circle) whose center is the pixel $(x_{1i}, y_{1i})$ and whose radius is $\alpha_{ix}$, and $\theta_{iy1} = f(\pi\alpha_{iy}^2)$, brought by the circle (in fact, the area of the circle) whose center is the pixel $(<x_{2i}>, y_{1i})$ and whose radius is $\alpha_{iy}$, i.e.,

$$\theta_{i1} = f(\pi\alpha_{ix}^2) \cdot f(\pi\alpha_{iy}^2)$$

$$f(\pi\alpha_{ix}^2) \cdot f(\pi\alpha_{iy}^2) = f(\pi\alpha_i^2)$$

$$f(\pi\alpha_{ix}^2) \cdot f(\pi\alpha_{iy}^2) = f(\pi(\alpha_{ix}^2 + \alpha_{iy}^2))$$

$$\ln f(\pi\alpha_{ix}^2) + \ln f(\pi\alpha_{iy}^2) = \ln f(\pi(\alpha_{ix}^2 + \alpha_{iy}^2))$$

On both sides, as to $\alpha_{ix}$, the partial derivatives are got. We have

$$\frac{2\pi\alpha_{ix} \cdot f'(\pi\alpha_{ix}^2)}{f(\pi\alpha_{ix}^2)} = \frac{2\pi\alpha_{ix} \cdot f'(\pi(\alpha_{ix}^2+\alpha_{iy}^2))}{f(\pi(\alpha_{ix}^2+\alpha_{iy}^2))}$$

$$\frac{f'(\pi\alpha_{ix}^2)}{f(\pi\alpha_{ix}^2)} = \frac{f'(\pi(\alpha_{ix}^2+\alpha_{iy}^2))}{f(\pi(\alpha_{ix}^2+\alpha_{iy}^2))}$$

Suppose that $F(x) = \dfrac{f'(\pi x)}{f(\pi x)}$, then

$$F(\alpha_{ix}^2) = F(\alpha_{ix}^2+\alpha_{iy}^2) = F(\alpha_i^2) = D_1, \quad \text{where } D_1 \text{ is constant}$$

$$\frac{f'(\pi\alpha_i^2)}{f(\pi\alpha_i^2)} = D_1$$

$$2\pi\alpha_i \cdot \frac{f'(\pi\alpha_i^2)}{f(\pi\alpha_i^2)} = D_1 \cdot 2\pi\alpha_i$$

$$\frac{d\ln f(\pi\alpha_i^2)}{d\alpha_i} = \frac{d(D_1 \cdot \pi\alpha_i^2)}{d\alpha_i}$$

$$\ln f(\pi\alpha_i^2) = D_1 \cdot \pi\alpha_i^2 + D_2, \quad \text{where } D_2 \text{ is constant}$$

$$f(\pi\alpha_i^2) = D_3 \cdot e^{D_1 \cdot \pi\alpha_i^2}, \quad \text{where } D_3 \text{ is positive constant}$$

Besides, when $\alpha_i = 0$, obviously, $\theta_{i1} = f(\pi\alpha_i^2) = 1$

$$\therefore D_3 = 1$$

$$\therefore \theta_{i1} = e^{D_1 \cdot \pi\alpha_i^2}$$

Besides, the greater $\alpha_i$ is, the smaller $\theta_{i1}$ is; the smaller $\alpha_i$ is, the greater $\theta_{i1}$ is, so $D_1 < 0$. Suppose that $D_1 \cdot \pi = -C_1$, then $\theta_{i1} = e^{-C_1\alpha_i^2}$. It just conforms to normal distribution.

Then, we consider **a recognition accuracy-rate** $\theta_{i2}$ that is related to a colour recognition. It is not difficult to see that **the recognition accuracy-rate** $\theta_{i2}$ is closely related to the colour deviation of the two pixels $\beta_i$. Similar to $\theta_{i1}$

$$\theta_{i2} = f(C_2'\beta_i), \quad \text{where } C_2' \text{ is positive constant}$$

Similarly, **a recognition accuracy-rate** $\theta_{i3}$ that is related to a depth recognition

must conform to

$$\theta_{i3} = f(C_3' \beta_i'), \quad \text{where } C_3' \text{ is positive constant}$$

It is not difficult to see that **the recognition accuracy-rate** $\theta_i$ that the two pixels are matched with each other is the multiplication of $\theta_{i1}$, $\theta_{i2}$ and $\theta_{i3}$, i.e.,

$$\theta_i = \theta_{i1} \cdot \theta_{i2} \cdot \theta_{i3}$$

$$\theta_i = e^{-C_1 \alpha_i^2 + D_1 C_2' \beta_i + D_1 C_3' \beta_i'}$$

$$\theta_i = e^{-C_1 \alpha_i^2 - C_2 \beta_i - C_3 \beta_i'} \quad \text{where } C_1, C_2, C_3 \text{ is a positive constant}$$

In the following algorithm, an object of **a pattern classification tree** is **a contour feature** corresponding to a node in **the pattern classification tree.**

**Algorithm 3.5** Many objects of **a pattern classification tree** are recognized from an inputted scene.

At the recognition beginning, a root node of **a pattern classification tree** (an isolated node is included. The following are as such.) in the association cortex in the left hemisphere is at waking state. Through a corresponding neuron in corpus callosum, the nervous excitation activates **a contour feature** in the right hemisphere.

3-dimensional representation of an inputted scene is transformed into 2-dimensional representation by **Algorithm 2.9**. Then the 2-dimensional representation is transformed into 1-dimensional representation by **Algorithm 2.10**. Suppose that the edge-length of the square frame limiting (containing) the contours on the 1-dimensional representation is $n_2$, and the edge-length of the square frame limiting (containing) **the contour feature** corresponding to the root node is $w$. **Algorithm 2.11** is called again and again. Through the conversion of two adjacent **conversion planes**, the 1-dimensional representation (and the contained square frame) is similarly compressed to of the original size, in which $l \in N$, and $w \leq \dfrac{n_2}{2^l} < 2w$. Then, if necessary, **Algorithm 2.12** is successively called. The compressed 1-dimensional representation is successively enlarged. In other words, the 1-dimensional representation is similarly compressed to $\dfrac{1}{2^i}$ of the original size, in which $i = l-1$, $l-2 \ldots 1$, $0$, $i$ is successively decreased. Then excitation of neurons on **the contour feature** corresponding to the root node sends along vertical direction of **conversion planes,** wakens **a contour feature** of identical size on the plane where the compressed or not 1-dimensional representation is. Every position (left, right, up and down) in every compressed square frame is searched with the new **contour feature**. The neurons (at waking state) on **the contour feature** can be approximately synchronously matched with the neurons on the group of contours similar to **the contour feature** between the one multiple of **the contour feature** and its double multiple no matter how much multiple value (how much depth value) the group of contours is at and where it is in the corresponding square frame. At the same time, **Algorithm 3.1** is called again and again. And every corresponding **average recognition accuracy-rate** is got by **Algorithm 3.3** again and again. When a group of contours is matched with **the contour feature** corresponding to the root node at a spot in the corresponding square frame on a multiple value, if **an average recognition accuracy-rate** $\bar{\theta}$ is greater than or equal to the threshold $C$, the corresponding recognition is successful. Then **Algorithm 3.6** is called. And a single object of **the pattern classification tree** is successively (parallelly) recognized from the inputted scene.

Traversal is continued forward from the spot in the corresponding square frame on the multiple value. When **the contour feature** corresponding to the root node is recognized from the inputted scene again.**Algorithm 3.6** is called again and again. the other object of **the pattern classification tree** is successively recognized from the inputted scene.

The cycle is as such. At last, one object, several ones, or none of one of **the pattern classification tree,** is recognized from the inputted scene.

**Parallel Algorithm 3.6** A single object of **a pattern classification tree** is successively (or parallelly) recognized from an inputted scene.

At the recognition beginning, a root node of **a pattern classification tree** (an isolated node is included) in the association cortex in the left hemisphere is at waking state. the root node successively (or parallelly) activates its every child node. And every child node activates a corresponding **contour feature** in the right hemisphere.It is assumed that the recognition of the root node is successful at a center point $(x, y)$ on $2^{-k}$ ($k \in N \bigcup \{0\}$) compressed multiple of the 1-dimensional representation of the inputted scene. At the center point $(<0.5x>,<0.5y>)$, $(x, y)$ and $(2x, 2y)$ successively, on $2^{-k-1}$, $2^{-k}$ and $2^{-k+1}$ compressed multiple of the 1-dimensional representation of the inputted scene, correspondingly, **the contour feature** corresponding to every child node is successively (or parallelly) used to matching treatment, **Algorithm 3.1** is called again and again, and every corresponding **average recognition accuracy-rate** is got by **Definition 3.2** again and again (if $k=0$, the center point of $(2x, 2y)$ and $2^{-k+1}$ compressed multiple are not considered) (The 1-dimensional representation of the inputted scene is compressed to the half of the original size directly by the nodes activated just now on the above plane again. It is enlarged to the double of the original one by **Algorithm 2.12** again. It is noted that $1 \leq \mu < 2$, $\mu$ appears in **Algorithm 3.1**). If **an average recognition accuracy-rate** $\bar{\theta}$ is greater than or equal to the threshold $C$ during the matching of **the contour feature** corresponding to a child node, then the corresponding recognition of the child node is successful, otherwise, it is not successful. Every child node of unsuccessful recognition and its corresponding **contour feature** in the right hemisphere will be blocked, and will return to sleeping state. As to the sequential algorithm, if the recognition of a child node is successful, every child node after it need not recognition treatment. As such, the recognition treatment is conducted down **the pattern classification tree** until a node is a leaf one, or none of the recognition of its children nodes is successful. **The contour feature** corresponding to the node is required.

From the above algorithm, we know that **Algorithm 3.6** may be not only sequential, but also parallel.

If, in a memory (recognition) procedure, many patterns (nodes) can be searched in **a pattern classification forest** in the association cortex in the left hemisphere by an inputted scene, and these patterns are subjected to the same or different **pattern classification trees**, the following algorithm is used.

**Parallel Algorithm 3.7** Many objects of **a pattern classification forest** are successively (or parallelly) recognized from an inputted scene.

As to every **pattern classification tree** (every isolated node is included) in **the pattern classification forest** in the association cortex in the left hemisphere,**Alg.3.5** is successively (parallelly) called. At last, all nodes musted to the following conditions are searched in the **pattern classification forest:** a node is a leaf one, or none of the recognition of its children nodes is successful.These nodes may be subject to the same or different **pattern classification trees.** The **contour features** corresponding to these nodes are required.

From the above algorithm, we know that **Algorithm 3.7** may be not only sequential, but also parallel.**Algorithm 3.8 A procedure that a thing happens** is recognized from an inputted continuous scene sequence (At every time, some **contour units** may simultaneously appear).

As to every scene in the sequence, **Algorithm 3.7** is successively called. Every time, that is activated by all object neurons (recognized nodes) recognized just now, in time sequence recorded in hippocampus, that there are several or not, is regarded as a common time of these object neurons.

(1) On the one hand, from space, since these common times have been activated in hippocampus, they will also activate some neurons in the association cortex in the left hemisphere, respectively, and if the neurons activated by a common time are matched with all object neurons recognized just now, one to one,the activated state of the common time will be temporarily memorized, otherwise, not only will the neurons not matched be restrained themselves, to return to sleeping state again, but also will the corresponding common time be done;

(2) on the other hand, from time, these common times in hippocampus depend on subsequent times being continuously activated or not, respectively, and if subsequent times are continuously activated, excitable value of the corresponding common time can be maintained, otherwise, after a short time, excitable value of these common times will be gradually decreased until these common times return to sleeping state again.

If there is a segment of continuous time sequence in which every time is activated, and whose time span is the same as the one of an inputted continuous scene sequence, then the recognition is successful, the segment of time sequence represents a recognized **procedure that a thing happens**, otherwise, it is not successful.

Because **Algorithm 3.7** may be not only sequential, but also parallel, **Algorithm 3.8** may be as such.

## 4. Algorithms of Pattern Learning

By the method that a newly inputted pattern and an old one in the pattern-base are recognized from and matched with each other, the position, at which the new pattern may be put, may be found from the pattern-base memorized by category. then the new pattern is put at the position. In the introduction section, the detail has been described, so we omit it here.

The algorithm that **a contour feature** is recognized from a recognized **contour unit** is similar to the front section of **Algorithm 3.5**, i.e., similar to the situation that the first object of **a pattern classification tree** is recognized from an inputted scene.

**Algorithm 4.1 A contour feature** is recognized from a recognized **contour unit.**

The algorithm is divided into 2 cases, 2 cases are all similar to the front section of **Algorithm 3.5**:

(1) If the recognized **contour unit** is a newly learned **contour unit**, the newly learned **contour unit** is regarded as an inputted scene,and **the contour feature** corresponding to the current node in **a pattern classification forest** is regarded as the required **contour feature**. Its treatment method is similar to the front section of **Algorithm 3.5**.

(2) If the recognized **contour unit** is an old **contour unit** corresponding to the current node in **a pattern classification forest**, at first, every neuron on **the contour feature** corresponding to the node is activated,**a position feature value** permanently recorded in polypeptide contained in every neuron is also activated. Besides, by the treatment method similar to the one of an inputted scene in the front section of **Algorithm 3.5**, the learned **contour unit** is converted into the corresponding **contour feature** that is regarded as the required one. the required **contour feature** is used to be matched with the above **contour feature.** its treatment method is similar to the front section of **Algorithm 3.5**. If the recognition is not successful, **Algorithm 2.14** is successively called. Through the conversion of 2 adjacent **conversion planes**,the 1-dimensional representation of the above **contour feature** or its enlarged $2^k$ multiple successively activates the 1-dimensional representation of the enlarged double of the original size on the lower plane. Then the required **contour feature** that must be previously projected on the lower plane is successively used to be matched with the enlarged 1-dimensional representation. At last, the recognition is either successful, or unsuccessful.

**Algorithm 4.2** A direct connection, from a learned **contour feature** to an isolated neuron in the association cortex in the left hemisphere, is established. Then the direct connection, from the neuron to hippocampus, is established. Besides, **Position coordinates** of a corresponding **position neurons** in corpus callosum are permanently recorded in every neuron on the learned **contour feature** and its successively enlarged 1-dimensional representations.

Every pixel (neuron) on the learned **contour feature** and its successively enlarged 1-dimensional representation is at waking state. Every neuron on the learned **contour feature,** that is the closest to corpus callosum, and has been already compressed into the minimum, releases an unique chemical substance that is a hormone[8].Right now,the chemical substance prompts **the position neurons**, that are not linked to the neurons in the right hemisphere, in corpus callosum, to dynamically grow synapses towards every neuron on **the contour feature**.When a synapse of **a position neuron** in corpus callosum arrives at a neuron on **the contour feature**, **the position neuron** in corpus callosum interacts with every neuron on **the contour feature**, and the other unique chemical substance is released from every neuron on **the contour feature**, the chemical substance can promote the other synapses of **the position neuron** in corpus callosum to rapidly grow, at last, every neuron on **the contour feature** is arrived at by a synapse, but the chemical substance can inhibit the synapses of the other **position neuron** in corpus callosum to grow, shrivel back the synapses.

Then **the position neuron** in corpus callosum will dynamicly grow a synapse towards the association cortex in the left hemisphere until the synapse is linked to an isolated neuron. The chemical signal released from every isolated neuron ought to be different from the one done from every neuron previously linked to **a position neuron** in corpus callosum. These two chemical signals can help the synapse linked to an isolated neuron. The neuron represents the learned **contour feature** in the right hemisphere.

At last, the neuron, which the synapse is just linked to, will be activated, and be at waking state. under the help of the chemical signal released by **the present neuron** in hippocampus, it will dynamicly grow a synapse towards **the present neuron** in hippocampus until the synapse is linked to **the present neuron** in hippocampus.

Besides, **the position coordinates** of the corresponding **position neuron** in corpus callosum will be recorded in every neuron on the learned **contour feature**. They will be parallelly put together with **the feature value** (**the relative colour value** and **the relative depth value**) newly memorized in every neuron. **A position feature value** will be formed. It will be permanently recorded in polypeptide contained in every neuron. At the same time, **Algorithm 2.13** will be called, again and again. **The position coordinates** of the corresponding **position neurons** in corpus callosum, which will have just (permanently) been recorded in the neurons at waking state on the above plane, will be sent down, be permanently recorded in the neurons at waking state on the lower plane.

**Algorithm 4.3** A pattern is learned[5]

At the learning beginning, every root node (an isolated node is included) in **a pattern classification forest** in the association cortex in the left hemisphere is at waking state. Their excitation respectively activates corresponding **contour features** in the right hemisphere through corresponding **position neurons** in corpus callosum. Firstly, **a contour unit** corresponding to every root node (an isolated node is included) in **the pattern classification forest** is successively regarded as a recognized object. **Algorithm 4.1** is called again and again. The recognition process is conducted with the newly inputted **contour feature**:

(1) If the newly inputted **contour feature** can be recognized from a **contour unit** or more corresponding to a root node or more (The number is $l$) (Every **average accuracy-rate** $\bar{\theta}$ is greater than a threshold. The following are as such.):

(1.1) If $l \geq 2$, then, at first, **Algorithm 4.2** is called, the direct connection, from the learned **contour feature** to a neuron in the association cortex in the left hemisphere, is established, and the direct connection, from the neuron to hippocampus, is established, and the neuron (the new node) is regarded as the father one of the

$l$ root nodes, and an unique chemical substance that is a hormone is released from everyone of the $l$ root nodes (at waking state), the chemical substance can promote the new node (at waking state) to grow some synapses that respectively arrive at everyone of the $l$ root nodes. so a larger **pattern classification tree** is formed. **Algorithm 4.3** is ended.

(1.2) If $l = 1$, then **Algorithm 4.1** is called:

(1.2.1) If **the contour feature** corresponding to the root node can be recognized from the newly inputted **contour unit**, then **the contour feature** that the newly inputted **contour unit** is transformed into is the same as the one corresponding to the root node, and the root node is turned into the one, corresponding to the newly inputted **contour unit**, in the association cortex in the left hemisphere, and the newly inputted **contour unit** and the **contour feature** that it is transformed into are deleted (forgot). **Algorithm 4.3** is ended.

(1.2.2) Otherwise, similar to step (1.1), the new node is regarded as the father one of the root node. **Algorithm 4.3** is ended.

(2) If the newly inputted **contour feature** cannot be recognized from the **contour unit** corresponding to anyone of the root nodes, then the newly inputted **contour unit** must be regarded as a recognized object, and **Algorithm 4.1** is called again and again, and the recognition process is successively conducted with **the contour feature** corresponding to every root node:

(2.1) If none of the root nodes is successfully recognized, then, at first, **Algorithm 4.2** is called, the direct connection, from the learned **contour feature** to a neuron in the association cortex in the left hemisphere, is established, and the direct connection, from the neuron to hippocampus, is established, and the neuron (the new node) is regarded as an isolated node, i.e., the direct connection, from it to the other neuron in the association cortex in the left hemisphere, is not established. **Algorithm 4.3** is ended.

(2.2) If there are just or more than two root nodes that are successfully recognized, then, because nothing can simultaneously belong to two types of different things, at this time, the learning is error. **Algorithm 4.3** is ended.

(2.3) If there is a root node $d$ which is successfully recognized, the node $d$ must activate every child node, then every child node of $d$ must be treated with the method similar to every root node, at this time, there are several similar things as follows:

(a) Similar to step (1.1), then the new node is put into a pattern-base, and the new node is regarded as a child node of $d$, at the same time, it is regarded as a father one of several original children nodes of $d$, the synapse connections between these nodes are established. **Algorithm 4.3** is ended.

(b) Similar to step (1.2.1), then the newly inputted **contour feature** is the same as the one corresponding to a child node of $d$, and the child node is turned into the one, corresponding to the newly inputted **contour feature**, in the association cortex in the left hemisphere, the newly inputted **contour unit** and the **contour feature** that it is transformed into are deleted (forgot). **Algorithm 4.3** is ended.

(c) Similar to step (a) and step (1.2.2), then the new node is regarded as a father one of an original child node of $d$. **Algorithm 4.3** is ended.

(d) Similar to step (2.1), then the new node is put into the pattern-base, and the new node is regarded as a child node of $d$, the synapse connection between them is established. **Algorithm 4.3** is ended.

(e) Similar to step (2.2), then the learning is error. **Algorithm 4.3** is ended.

(f) Similar to step (2.3), then the new node is regarded as an offspring of a child node $d^{'}$ of $d$, if $d^{'}$ is a leaf node, and only the new node is directly appended, the synapse connection between them is established, otherwise, recursive process as such is continued, layer by layer. **Algorithm 4.3** is ended.

From the above algorithm, we know that **Algorithm 4.3** may be not only sequential, but also parallel.

**Algorithm 4.4 A procedure that a thing happens** is learned from an inputted continuous **contour unit** sequence (At every time, many **contour units** may simultaneously appear.).

At first, at every regular time interval $\Delta T$, a signal representing that time is produced from **the present neuron** in hippocampus and memorized in it. At the learning beginning, suppose that the time that was memorized in **the present neuron** in hippocampus at the last time is $T$. At every time $T+i*\Delta T$, in which $0 \leq i \leq I$, as to every **contour unit** simultaneously appearing in continuous sequence, the process as follows is conducted: At first, the $(i, j)$-th **contour unit** represents the $j$-th **contour unit** appearing at the time $T+i*\Delta T$, in which $0 \leq j \leq J_i$. As to every one in the $(i, j)$-th **contour units**, in which $0 \leq j \leq J_i$, **Algorithm 4.3** is successively (parallelly) called. Every neuron, corresponding to these **contour units**, in the association cortex in the left hemisphere, is at waking state. Not only are these neurons just linked to these **contour units,** but also are they in **a pattern classification forest**, previously. A signal representing a new time $T+i*\Delta T$ is produced from **the present neuron** in hippocampus and memorized in it. Besides, the signal representing the time $T+i*\Delta T$ is sent down along the linear structure composed of **time neurons** in hippocampus.When it is sent to the synapses of these neurons at waking state, it is sent to these neurons by the synapses, and recorded in multipeptide contained in these neurons. As such, **a procedure that a thing happens** $<T,T+I*\Delta T>$ is memorized in a human brain.A segment of time sequence $<T,T+\Delta T>,<T,T+2\Delta T>…<T,T+I*\Delta T>$ is recorded in **time neurons** in hippocampus.

From the above algorithm, we know that, if **a contour unit** appears many times,signals representing many times must be recorded in multipeptide contained in a neuron, in the association cortex in the left hemisphere in a human brain, corresponding to **the contour unit.**

From the above algorithm, we know that **Algorithm 4.4** may be not only sequential, but also parallel.

## 5. Summary and Outlook

In this paper, we study the following pattern recognition problem: Every pattern is a 3-dimensional graph, its surface can be split up into some regions, every region is composed of the pixels with the approximately same color value and the approximately same depth value that is distance to eyes, and there may be also some contours, e.g., literal contours, on the surface of every pattern. In this paper, we propose the cognitive model aiming at the recognition of the type of patterns. After patterns in a specific area are learned, a pattern-base stored by classification in the area may be built in a human brain.Coupled with pattern recognition function, a scene in the specific area may be recognized in the brain. In [1], a coarser model or a basicer one is described. In this paper, some important errors are revised, some key things are added, at last, a complete model is described.

This type of cognitive model plays a major role in visual recognition.The main cognitive model of auditory recognition and the main cognitive model of somatosensory recognition are sometimes used to visual recognition. But the later two types of cognitive model only are auxiliary to visual recognition. Besides, The main cognitive model of visual recognition is sometimes used to auditory recognition and somatosensory recognition. But it is only auxiliary. By the way, when the main cognitive model of visual recognition is used to somatosensory recognition, several types of colour values are replaced by values of temperature, touching, pain and pressuring, separately, and depth recognition is invalid. The fuzzy and not accurate characteristics on the above two hands are inherent and insurmountable.

The main cognitive model of auditory recognition and the main cognitive model of somatosensory recognition will be described in other papers.

## 6. Thanks

Sincerely dedicate this paper to my mentor, Mr. Lu Ru-Qian, an academician. I thank him for his meticulous concern and earnest teaching for me.

## 7. Reference

[1] Chen Yonghong, The Cognitive Model towards Contour Recognition

[2] Chen Yonghong, The Main Cognitive Model of Auditory Recognition:

   The Recognition of Characteristic Values of a Signal

[3] Chen Yonghong The Main Cognitive Model of Somatosensory Recognition:

   Recognition of Characteristic Pixels of an Image

[4] Diyun Ruan, Tiande shou, Neuro physiology, Published by University of Science and Technology of China, p276-298.

[5] Lu Ruqian, Artificial Intelligence, second volume, Publisher of Science, part of decision tree

[6] George Adelman edit, Encyclopedia of Neuroscience, Corpus Callosum

[7] George Adelman edit, Encyclopedia of Neuroscience, Hippocampus

[8] George Adelman edit, Encyclopedia of Neuroscience, Memory, Hormone Intluences